\documentclass[useAMS,usenatbib]{mnras}

\usepackage[utf8]{inputenc}
\usepackage{graphicx}   
\usepackage{amsmath}    
\usepackage{amssymb}    
\usepackage{multicol}   
\usepackage{bm}         
\usepackage{pdflscape}  
\usepackage{float}      
\usepackage{tipa}       
\usepackage{orcidlink}
\usepackage{color, soul}
\soulregister\cite7
\soulregister\ref7
\soulregister\pageref7

\hypersetup{final}
\title[]{Simulating stellar coronal rain and slingshot prominences}
\author[]{Daley-Yates S.}

\author[Simon Daley-Yates \& Moira M. Jardine]{
    S. Daley-Yates$^{1}$\thanks{E-mail: sddy1@st-andrews.ac.uk} \orcidlink{0000-0002-0461-3029},
    Moira M. Jardine$^{1}$ \orcidlink{0000-0002-1466-5236}\\   
    $^{1}$School of Physics and Astronomy, University of St Andrews, North Haugh, St Andrews, Fife, Scotland KY16 YSS, UK
}

\begin{document}

\date{}

\pagerange{\pageref{firstpage}--\pageref{lastpage}} \pubyear{2023}

\maketitle

\label{firstpage}

\begin{abstract}
We have numerically demonstrated that simulated cool star coronae naturally form condensations. If the star rotates slowly,  with a co-rotation radius greater than the Alfv\'{e}n radius (i.e.  $R_{\mathrm{K}} > R_{\mathrm{A}}$), these condensations will form below the co-rotation radius $R_{\mathrm{K}}$ and simply fall back to the stellar surface as coronal rain.  If, however, the star is more rapidly rotating, ($R_{\mathrm{K}} < R_{\mathrm{A}}$), not only rain will form but also ``slingshot prominences''. In this case, condensations collect into a large mass reservoir around the co-rotation radius, from which periodic centrifugal ejections occur. In this case, some $51\%$ of the coronal mass is cold gas, either in rain or prominences.  We find that 21\% of the mass lost by our simulated fast rotating star is cold gas. Studies of stellar mass-loss from the hot wind do not consider this component of the wind and therefore systematically underestimate mass-loss rates of these stars. Centrifugal ejections happen periodically, between every 7.5 - 17.5 hours with masses clustering around $10^{16}$ g, These results agree well with observational statistics. Contrasting the fast and slow rotating magnetospheres, we find that there are two distinct types of solutions, high lying and low lying loops. Low lying loops only produce coronal rain whereas high lying loops produce both rain and slingshots. 
\end{abstract}


\section{Introduction}

Much of our understanding of the structure and dynamics of stellar coronae has come from X-ray or UV observations of the hot (1-10 $\times 10^6$K) gas trapped in the star's corona by the confining effect of the magnetic field. The hot plasma that escapes to form the stellar wind is not dense enough to be observed directly and is mainly studied through its impact on the rotational evolution of the star. The mechanisms responsible for heating the coronal gas to these temperatures are still a matter of debate, many decades after the existence of stellar coronae was discovered. Studies of both this background heating and also the intermittent powerful heating in stellar flares have however been re-invigorated by the realisation of their impact on the evolution of exoplanetary atmospheres \citep{2005AsBio...5..706S,2020IJAsB..19..136A}.

Observations of the Sun, however, show that coronal gas can also exist in a {\it cool} ($~10^4$K) phase. These are temperatures typical of the Sun's chromosphere, but they are found within the solar corona both in large, quasi-stable prominences (masses typically $10^{15}$ g) and also in smaller, dynamic clumps of ``coronal rain'' \citep{2022FrASS...920116A,2023arXiv230508775S}. This rain is often found to be falling at speeds close to, but typically less than, the free-fall speed, with values in the range $100-150  \ \mathrm{km} \ \mathrm{s}^{-1}$  \citep{2022FrASS...920116A}, or $\simeq 40 \ \mathrm{km} \ \mathrm{s}^{-1}$ \citep{2023arXiv230508775S}. Sun-as-a-star measurements give similar values. \citet{2022ApJ...933..209N} found $95 \ \mathrm{km} \ \mathrm{s}^{-1}$  while  \citet{2022ApJ...939...98O} reported velocities up to $200 \ \mathrm{km} \ \mathrm{s}^{-1}$.

This is consistent with the view that these clumps are condensations that are falling along magnetic field lines. They are often found after a flare has ablated chromospheric material which rises into the corona, raising the local density to the point where it cools and then falls back towards the surface. This cycle of heating and cooling regulates the total mass of the corona - indeed above some active regions the cool gas can comprise some $~50\%$ of the active region volume \citep{2022ApJ...931L..27S}.

Studies of this cool gas phase in the coronae of other stars originally focused on the very large, relatively stable ``slingshot'' prominences \citep{collier1989I,collier1989II}. These cool clouds are detected as transient absorption features that move from blue to red through the H$\alpha$ line profile. They often recur at the same rotation phase, suggesting that they are co-rotating with the star. The time taken for the absorption feature to move through the line profile gives the distance of the absorbing material from the rotation axis directly, showing that they typically form close to or beyond the Keplerian co-rotation radius, where the outward centrifugal force begins to dominate over the inward gravitational force. Since their original discovery, they have been detected in a range of stars, from those still in the T Tauri phase \citep{Skelly2008,Skelly2009} to those whose disks have dissipated, but which are still rotating rapidly  \citep{collier1992,Hall1992,Byrne1996,Eibe1998,Barnes2000,Barnes2001,Petit2005,Dunstone2006,DunstoneThesis2008,2016MNRAS.463..965L,2020A&A...643A..39C,2021A&A...654A..42C,2021MNRAS.504.1969Z}. They have even been observed on binary stars {such as the K supergiant 32 Cyg \citep{1983A&A...124L..16S}}.  

The prominence masses derived from these observations are typically ($2-6 \times 10^{17}$ g). This is several orders of magnitude greater than large solar prominences, but consistent with predictions of theoretical models that use stellar surface magnetic maps as inputs \citep{2018MNRAS.475L..25V,waugh2019,2019MNRAS.485.1448V,waugh2021}. When these prominences become unstable, they will be ejected (if they are beyond the co-rotation radius)  and so may remove enough angular momentum from the star to contribute significantly to the wind torques \citep{2022MNRAS.513.5611F,waugh2022}. This suggests that prominence ejection may provide a significant fraction of the extra torque required for stellar wind models to explain the observed rotational evolution 
\citep{2013A&A...556A..36G,2015A&A...577A..98G,2018MNRAS.474..536S,2024MNRAS.529L.140E}.

The large distance from the stellar surface at which these ``slingshot'' prominences are supported means that they are expected to form and be ejected in a cyclic process \citep{2020MNRAS.491.4076J}. This is because the sonic point of the upflow that forms the prominence is typically below the prominence formation site (close to the co-rotation radius). As a result, the growth in mass of the prominence can not be checked by information propagating back down to the surface,  leading to runaway growth and eventual centrifugal ejection. The continued upflow ensures that another prominence forms and this limit-cycle behaviour continues. The observationally-derived masses and lifetimes of prominences therefore provide crucial information on the mass-loss rates of these stars. This is particularly important for very active stars where other methods have so far been unable to provide measurements of wind mass-loss rates \citep{2021ApJ...915...37W}. 

In addition to these large and quasi-stable structures, cool coronal plasma has also been detected in smaller-scale, more dynamic structures. Transient, red-shifted absorption features were detected in the earliest studies \citep{1987LNP...291..491B,1993A&A...274..245H,1999A&A...341..527E}. These were interpreted as ``failed prominences'' - material that had condensed, but failed to find a stable location in which to accumulate and so simply drained back to the surface.  More recent surveys have now shown that these transient features can be seen both shifted to the red and also to the blue \citep{2018A&A...615A..14F,Vida2019,2022ApJ...926L...5N}, with estimated masses in the range $10^{13}-10^{14}$kg \citep{Vida2019}. 

Typically, these features are moving at speeds below the escape speed, although some have been seen  at greater speeds that indicate that they may be undergoing ejection from the star. These fast-moving features have in many cases been observed to occur at the same time as extremely powerful ``superflares'' detected with Lamost \citep{2022ApJ...925..155K,2022ApJ...928..180W} or TESS \citep{2022NatAs...6..241N,Inoue2023,2023ApJ...945...61N,2024ApJ...961...23N}. The possibility that these ejections are associated with the stellar equivalent of solar coronal mass ejections  makes their study even more important. Solar coronal mass ejections are strong sources of energetic particles that can ionise the upper layers of planetary atmospheres. Their frequency and power on younger solar analogues, or exoplanet hosts, may be an important aspect of exoplanetary evolution.

Originally, theoretical studies of this ``coronal rain'' were confined to the solar case \citep{Antiochos1999, Karpen2006, Antolin2010, Froment2018, Li2022} with many focused on the role of the heating mechanism \cite{Xia2012, Fang2013, Zhou2021, 2021ApJ...920L..15R}. Recently, we published a study in which the concept of coronal rain is extended to the stellar case \citep{Daley-Yates2023}. By simulating a moderately-rotating solar-like star, we demonstrated the formation and subsequent draining of large-scale coronal condensations \citep{Daley-Yates2023}. These condensations were triggered by enhanced footpoint heating. The resultant coronal rain had line-of-sight velocities in the range $50 \ \mathrm{km} \ \mathrm{s}^{-1}$ (blue shifted) to $250 \ \mathrm{km} \ \mathrm{s}^{-1}$ (red shifted), typical of those inferred from stellar H$\alpha$ line asymmetries. Since this star was only rotating moderately quickly, its co-rotation radius lay beyond its corona. As a result, although these condensations were formed at all heights within the corona, condensations could not be formed beyond the co-rotation radius, and hence could not escape the star.

In the present paper we extend this pilot study to a more rapidly-rotating star to examine the outward ejection of such condensations; for the first time numerically demonstrating the slingshot mechanism. We also include an updated version of the slower rotation case described above so that we can contrast the two rotation states. Therefore we can see how faster rotators exhibit the slingshot mechanism and slower ones do not, within the same numerical prescription.

\section{Modelling}

For the basic equations solved in our simulations including the form of the magnetohydrodynamic (MHD) equations, please see our previous paper \citep{Daley-Yates2023} and the paper for the simulation code we use, \textit{MPI-AMRVAC} \citep{Keppens2021}.

In the following sections we will introduce the key concepts of the cooling instability, a phenomenological heating prescription and the magnetosphere classification needed to understand the results presented in Section \ref{sec:results}. We also describe how the magnetic field depends on rotation and therefore how the heating rate and magnetic field can be specified by the star's rotation rate.

\subsection{Stellar multi-phase gas}

The magnitude of radiative energy loss in a plasma depends on its temperature and composition, represented by a loss function $\Lambda (T)$. Between chromospheric and coronal temperatures ($10^{4} - 10^{6}$ K), $\Lambda (T)$ is non-linear, with cooler temperatures radiating away energy more efficiently. This leads to a runaway effect where coronal gas can suddenly cool, condensing to either long-lived prominences or short-lived coronal rain. It is the hot coronal plasma co-existing with the cooler condensations that we call a multi-phase gas. For a comprehensive study of the cooling function, the cooling instability and its role in numerical simulations see \cite{Hermans2021}.

\subsection{Magnetosphere classification}

Here we introduce a classification scheme where a star's magnetosphere is described as either dynamical (DM) or centrifugal (CM) based on whether the co-rotation radius ($R_{\mathrm{K}}$) is inside the Alfv\'{e}n radius ($R_{\mathrm{A}}$) or vice versa. This scheme is based on MHD simulations of massive stars by \cite{ud-Doula2002, ud-doula2006, ud-Doula2008}, see also the work of \cite{Petit2013}. {Indeed, the concept of centrifugal outbreak events that may be responsible for X-ray flares in massive stars \citep{ud-doula2006} may be very similar to slingshot prominence ejection in cool stars}. This notion was extended to cool stars by \cite{Villarreal2017} who postulated that both cool stars and massive stars trap material in their magnetospheres by centrifugal support in the same manner, despite having completely different underlying wind driving physics. Cool star winds are believed to be thermally driven \citep{parker1958} while massive star winds are believed to be radiatively driven \citep{CAK1975}. 

If $R_{\mathrm{K}} > R_{\mathrm{A}}$ then the star is a relatively slow rotator. All the magnetic loops have summits below the co-rotation radius, so that at all points along their length the effective gravity points downwards towards the stellar surface. Plasma that condenses in one of these loops will simply fall under the action of this effective gravity, back towards the stellar surface. If $R_{\mathrm{K}} < R_{\mathrm{A}}$ then the star is a relatively fast rotator. The summits of the tallest loops may be beyond the co-rotation radius. In this case, plasma that condenses there will tend to fall outwards but may be supported against centrifugal ejection by the tension of the magnetic field. Thus regions exist between $R_{\mathrm{K}}$ and $R_{\mathrm{A}}$ where there is mechanical stability and where plasma can accumulate and form a stable prominence. See the work of \cite{waugh2019, waugh2021, waugh2022} for a comprehensive analytic study of prominence stability and how it impacts cool star mass-loss.

To explore these concepts, we have simulated two stars: one with a centrifugal magnetosphere ($P_{\ast}~=~0.38$ days) and one with a dynamical magnetosphere ($P_{\ast}~=~3.8$ days) \footnote{We chose the value of $P_{\ast}~=~0.38$ days as it is the rotation rate of the K3 dwarf star BO Microscopii (BO Mic, HD 197890), AKA \textit{Speedy Mic}, a quintessential rapid rotating cool star with $R_{\mathrm{K}} < R_{\mathrm{A}}$ (giving a centrifugal magnetosphere). The choice of $P_{\ast}~=~3.8$ days is simply an order of magnitude lower so that $R_{\mathrm{K}} > R_{\mathrm{A}}$ (giving a dynamical magnetosphere).}. We interpret our results in Section \ref{sec:results} through this dynamical vs centrifugal magnetosphere classification scheme.

\begin{figure}
	\centering
	\includegraphics[width=0.5\textwidth,trim={0cm, 0cm, 0cm, 0cm},clip]{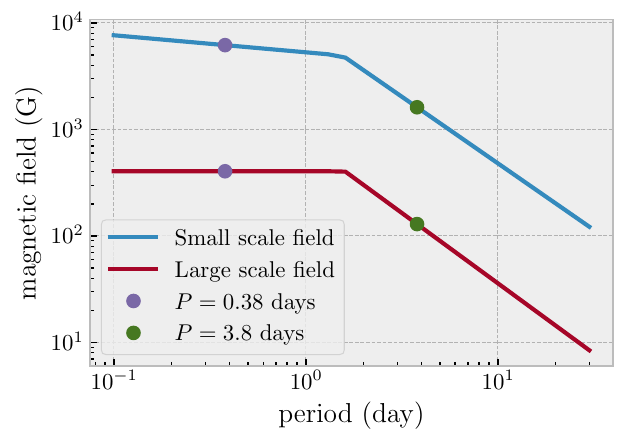}
	\caption[]{Scaling laws for magnetic field strength as a function of rotation period for cool stars. These are normalised such that the stellar radius and mass have solar values for all periods. These relations are from \cite{2022A&A...662A..41R} in the case of the small-scale field and \cite{Vidotto2014} in the case of the large-scale field. The {purple} dots indicate the field values for $P_{\ast}~=~0.38$ days and the {green} dots a value of $P_{\ast}~=~3.8$ days, corresponding to the values used in our two simulations.\label{fig:scaling_laws}}
\end{figure}

\begin{table*}
	\centering
	\caption[]{Stellar simulation parameters for the two magnetosphere types, centrifugal (CM) and dynamic (DM).
	\label{tab:parameters}}
	\begin{tabular}{cccc}
		\hline
		Name & Parameter & Centrifugal & Dynamic \\
		\hline
 		Radius & $R_{\ast}$ & 1 $R_{\sun}$ & 1 $R_{\sun}$ \\
		Mass & $M_{\ast}$ & 1 $M_{\sun}$ & 1 $M_{\sun}$ \\
		Period & $P_{\ast}$ & 0.38 day & 3.8 day \\
        Kepler radius (at equator) & $R_{\mathrm{K}}$ & 2.21 $R_{\ast}$ & 10.25 $R_{\ast}$ \\
        Polar magnetic field strength & $B_{0}$ & 810.0 G & 258.5 G \\
		Surface heating length-scale & $\lambda_{s}$ & 40 Mm & 40 Mm \\
		  Wind heating length-scale & $\lambda_{\rm{w}}$ & 2.14 $R_{\ast}$ & 1.03 $R_{\ast}$ \\
		  Surface heating amplitude & $H_{s}$ & $1.468 \times 10^{-1} \ \mathrm{erg} \ \mathrm{cm}^{-3} \ \mathrm{s}^{-1}$ & $2.543 \times 10^{-2} \ \mathrm{erg} \ \mathrm{cm}^{-3} \ \mathrm{s}^{-1}$ \\
		  Wind heating amplitude & $H_{\rm{w}}$ & $1.534 \times 10^{-3} \ \mathrm{erg} \ \mathrm{cm}^{-3} \ \mathrm{s}^{-1}$ & $5.281 \times 10^{-4} \ \mathrm{erg} \ \mathrm{cm}^{-3} \ \mathrm{s}^{-1}$ \\
		\hline
	\end{tabular}
\end{table*}

\subsection{Stellar wind heating model}

The following sections detail the specifics of our heating model and how we parameterise it with observational statistics of cool star magnetic fields. We show how the strength of the field, and therefore the heating, depends on rotation rate.

\subsubsection{Scaling laws}

We can measure the magnetic field strengths of stars using the Zeeman-Doppler Imaging (ZDI) and Zeeman Broadening (ZB) techniques (please see \cite{Donati2009} and references therein for details). We note that ZDI is sensitive to a star's large-scale magnetic field while ZB shows the small-scale magnetic field.

{In all the following equations we include the saturation of the magnetic field at periods below $P_{\mathrm{sat}} = 1.6$ days in accordence with the data from \cite{Vidotto2014} and \cite{2022A&A...662A..41R}.} By small-scale fields we mean magnetically active regions on the surface and for these we use the scaling relations of \cite{2022A&A...662A..41R}:
\begin{equation}
    B_{\mathrm{s}} = 8570 \ \mathrm{G} \ P^{-1.25} \ (P>P_{\mathrm{sat}})
    \label{eq:Reiners_scaling}
\end{equation}
\begin{equation}
    B_{\mathrm{s}} = 5300 \ \mathrm{G} \ P^{-0.16} \ (P<P_{\mathrm{sat}}).
\end{equation}
These are simplified versions of equations 2 and 3 in \cite{2022A&A...662A..41R}, we have removed the dependence on $M_{\ast}$, $L_{\ast}$ and $R_{\ast}$ \footnote{This is because we have simulated stars with only solar values for $M_{\ast}$, $L_{\ast}$ and $R_{\ast}$ and they appear in \cite{2022A&A...662A..41R} in solar units, therefore $M_{\ast}~=~1$, $L_{\ast}~=~1$ and $R_{\ast}~=~1$.} so that we assume that the magnetic field strengths depend only on rotational period.

For large scale fields, which here we mean the dipole component, we use the scaling relations of \cite{2007A&A...463...11H},
\begin{equation}
    B_{\mathrm{w}} = B_{\odot} \ \left( \frac{P}{P_{\odot}} \right)^{-1.32} \ (P>P_{\mathrm{sat}})
\end{equation}
\begin{equation}
    B_{\mathrm{w}} = B_{\odot} \ \left( \frac{P_{\mathrm{sat}}}{P_{\odot}} \right)^{-1.32} \ (P<P_{\mathrm{sat}}).
    \label{eq:Holzwarth_scaling}
\end{equation}
Unlike \cite{2007A&A...463...11H} we use the exponent of \cite{Vidotto2014} and account for magnetic field saturation. $B_{\odot}$ is the solar {polar} dipole field strength $\sim10$ G {at solar maximum \citep{Vidotto2014}}. These scaling relations and the values used in the simulations are plotted in Fig. \ref{fig:scaling_laws}.

\subsubsection{heating model}

The coronae and winds of our simulated stars are driven by a phenomenological heating model similar to that of \cite{Lionello2009} and \cite{Downs2010}. The following equation was derived by \cite{Abbett2007} and describes an empirical relation between the unsigned magnetic flux and energy deposited in the corona (see Fig 1. of \cite{Pevtsov2003} for this relation). The heating rate is given by
\begin{equation}
    Q = \frac{c \phi^{\alpha} \psi}{\zeta \int{\psi} dV}.
\end{equation}
Where $c~=~0.8940$ and $\alpha~=~1.1488$ are fit parameters from \cite{Bercik2005}. $\phi$ is the unsigned magnetic flux, given by
\begin{equation}
    \phi = \oint_{r=R_{\ast}} |B_{r}| ds
\end{equation}
and $\psi$ is the local heating weighting function which we simply take as the magnitude of the magnetic field as
\begin{equation}
    \psi = |\bm{B}| \exp \left( -\frac{r - R_{\ast}}{\lambda} \right).
\end{equation}
We also include an exponential envelope function limiting the heating to the lower corona, in the same way as \cite{Downs2010}. Because the volume integral of the weighting function is done over the entire simulation, this envelope function ensures the radial extent of the numerical grid does not impact the heating rate. 

The length scale of the envelope function has different values for the small-scale and large-scale heating. For the small-scale heating it is limited to 40 Mm, which is approximately the height of active regions.

When performing test simulations we found that for fixed size, large-scale heating envelopes, the wind from the star ($r~>~10 \ R_{\ast}$) would not maintain its temperature and would drop to chromospheric values. To address this, we use a scaling law based on rotation rate, that allows the heating to extend higher into the corona for faster rotating stars. Since faster rotating stars have stronger magnetic fields and therefore larger closed magnetospheres, this results in heating at greater altitudes. This scaling law is
\begin{equation}
    \lambda_{\mathrm{w}} = \lambda_{0} \left( \frac{P}{P_{\odot}} \right)^{-2} \ (P>P_{\mathrm{sat}})
\end{equation}
\begin{equation}
    \lambda_{\mathrm{w}} = \lambda_{0} \left( \frac{P_{\mathrm{sat}}}{P_{\odot}} \right)^{-0.4} \ (P<P_{\mathrm{sat}}).
\end{equation}
The exponents were determined experimentally, from trial-and-error simulations. 

The heating equations were applied to both the small- and large-scale fields with the resulting total heating being simply $Q_{\rm{total}}~=~Q_{\mathrm{s}}~+~Q_{\mathrm{w}}$, where the subscripts s and w refer to the heating rates derived from the small-scale (surface) and large-scale (wind) magnetic field strengths from equations \ref{eq:Reiners_scaling} and \ref{eq:Holzwarth_scaling} respectively.

All the variables used for modelling the heating and defining the physical parameters of our simulated stars are summarized in Table.~\ref{tab:parameters}.

\section{Numerical modelling}

As the simulations we present here are based on those in our previous study, we provide only a brief description of the simulation setup and highlight what we have changed for the purpose of the new simulations. For more details please see \citet{Daley-Yates2023}.

We solve the MHD equations with optically thin radiative losses and thermal conduction using the parallel, block based, adaptive mesh code MPI-AMRVAC \citep{Xia2018,Keppens2021}. Our computational grid extends between $r~\in~\{1, 50\} \ R_{\ast}$ and $\theta~\in~\{0, \pi\} \ \mathrm{radians}$. We used a resolution of 1024 cells in the radial direction and 640 cells in the poloidal direction. This differs from our previous study, in which we achieved an effective resolution of approximately twice this value via the use of additional refinement levels. In our present study we chose to use a static grid as we found in testing that the numerical overheads of adaptive mesh refinement outweighed the gains in resolution. This is because we needed high resolution both far from the star to resolve the slingshot prominences, as well as close to the star to capture the condensation behaviour there (in our previous study only the latter condensations were present).

In the next section we will present and discuss the results of our two simulations conducted with the framework described above.

\section{Results and Discussion}
\label{sec:results}

Here we present the results of our simulations, broken down into dynamical and centrifugal magnetospheres and describe the major phases seen in both. We will also show in the case of the centrifugal magnetosphere how the gas is partitioned into hot and cold phases, how the maximum temperature responds to centrifugal breakout, the impact this has on the mass-loss rate and finally the statistics of the simulated slingshots. 

\subsection{Magnetosphere types}

\begin{figure*}
	\centering
    \includegraphics[width=\textwidth,trim={0.9cm, 0.9cm, 0.9cm, 0.4cm},clip]{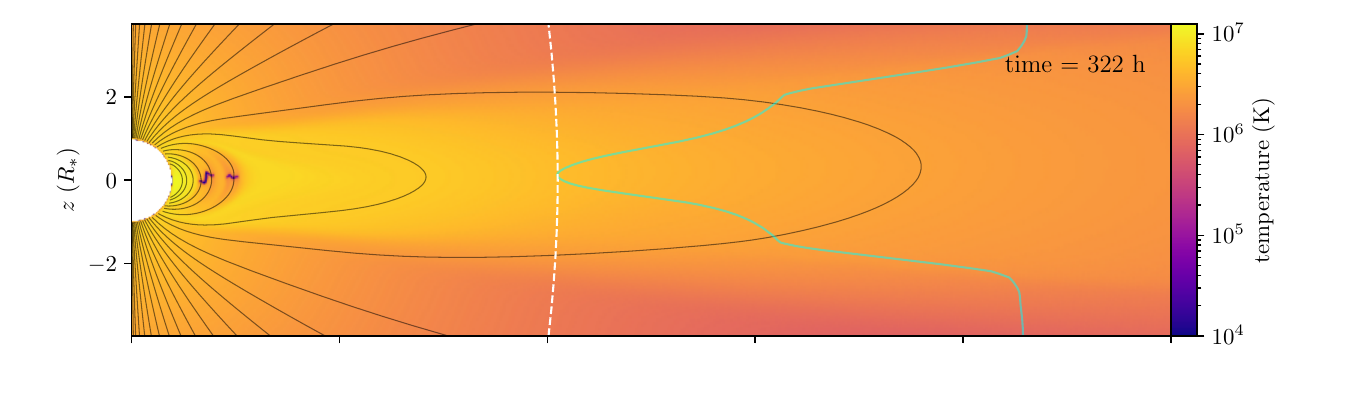}
    \includegraphics[width=\textwidth,trim={0.9cm, 0.9cm, 0.9cm, 0.4cm},clip]{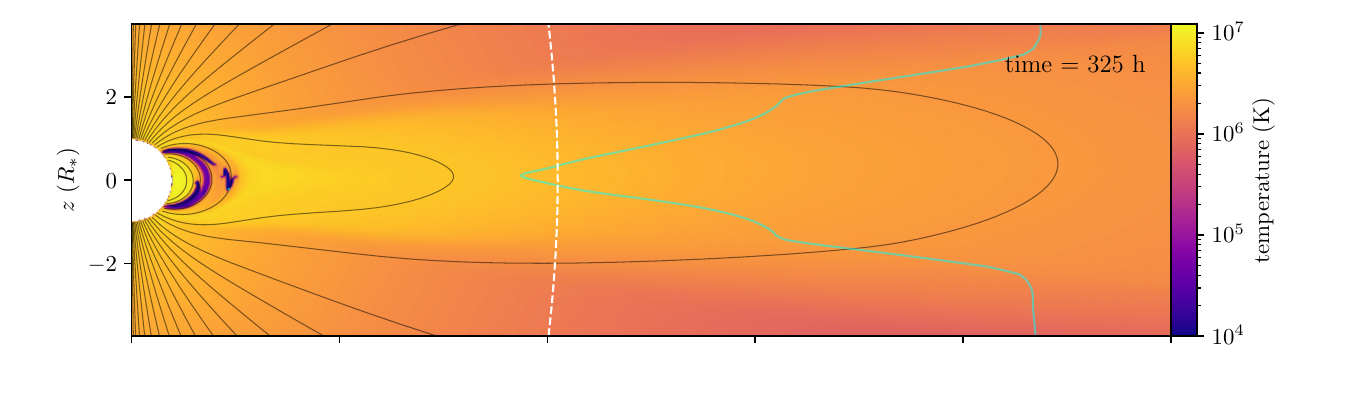}
    \includegraphics[width=\textwidth,trim={0.9cm, 0.9cm, 0.9cm, 0.4cm},clip]{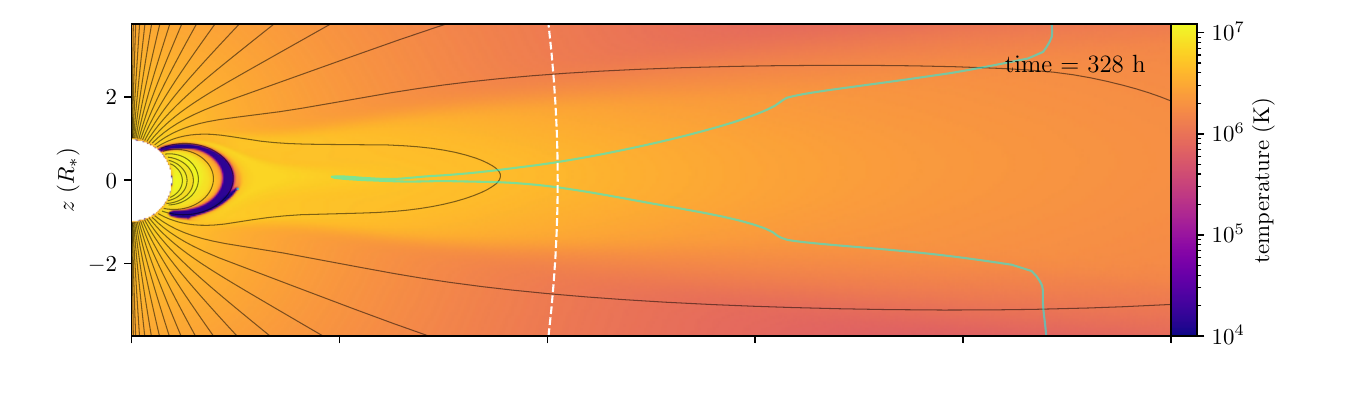}
    \includegraphics[width=\textwidth,trim={0.85cm, 0cm, 0.95cm, 0.3cm},clip]{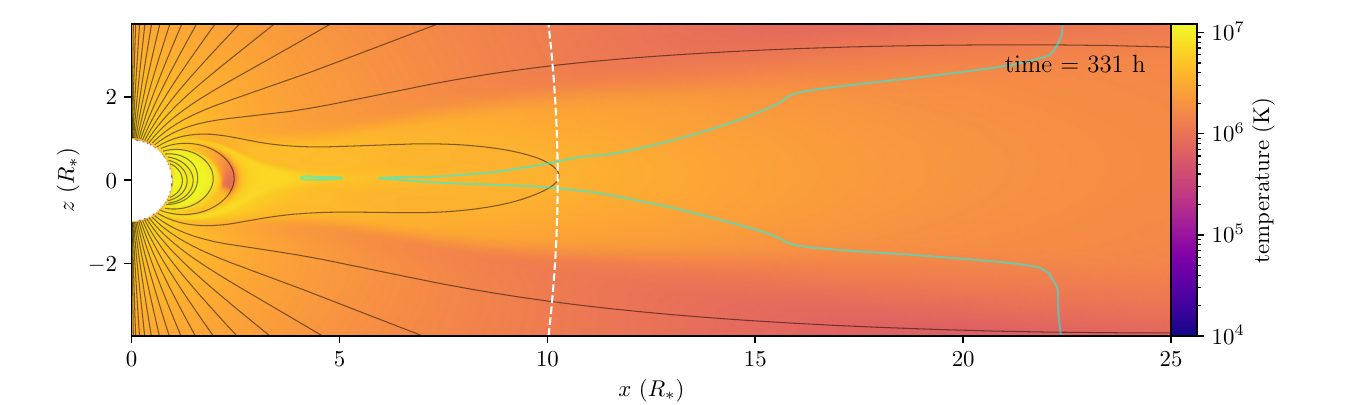}
	\caption[]{Time series of the 2D temperature profile for the dynamical magnetosphere case. There are three distinct phases of prominence formation: First, the thermal instability leads to gas cooling to chromospheric temperatures within the closed magnetosphere. Second, cold gas below $R_{\rm{K}}$ drains to the surface. Third, once all the cold gas has drained to the surface, there is only hot gas left in the magnetosphere. The dotted white line indicates the Kepler co-rotation radius and the cyan solid line shows the Alfv\'{e}nic Mach surface. A movie version of this figure is available online.
 \label{fig:evolution_dyn}}
\end{figure*}

\begin{figure*}
	\centering
    \includegraphics[width=\textwidth,trim={0.9cm, 0.75cm, 0.9cm, 0.5cm},clip]{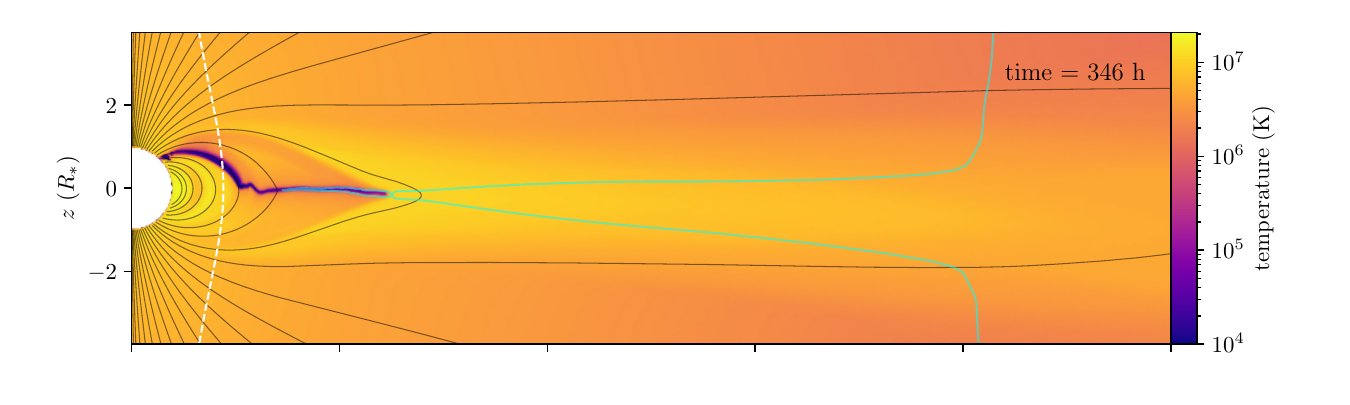}
    \includegraphics[width=\textwidth,trim={0.9cm, 0.75cm, 0.9cm, 0.5cm},clip]{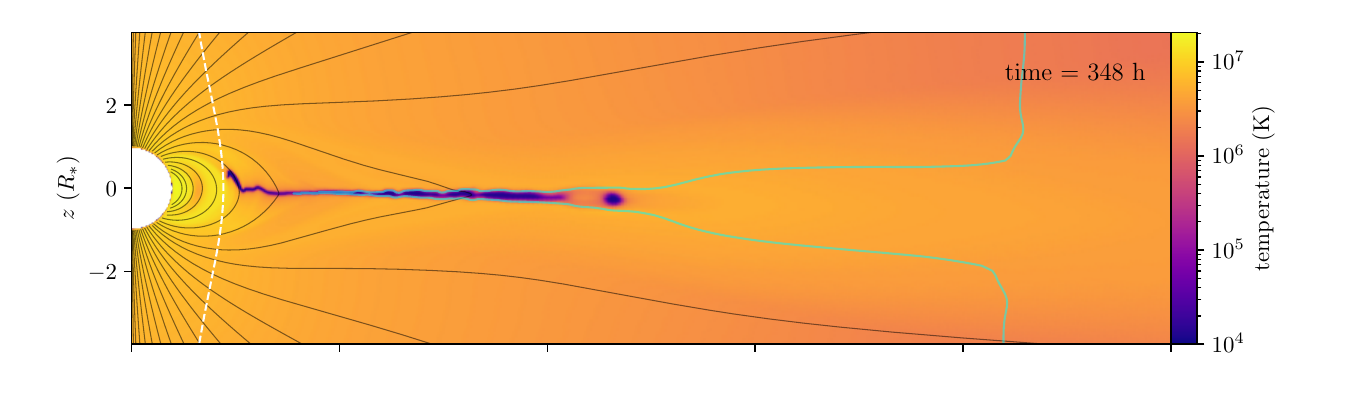}
    \includegraphics[width=\textwidth,trim={0.9cm, 0.75cm, 0.9cm, 0.5cm},clip]{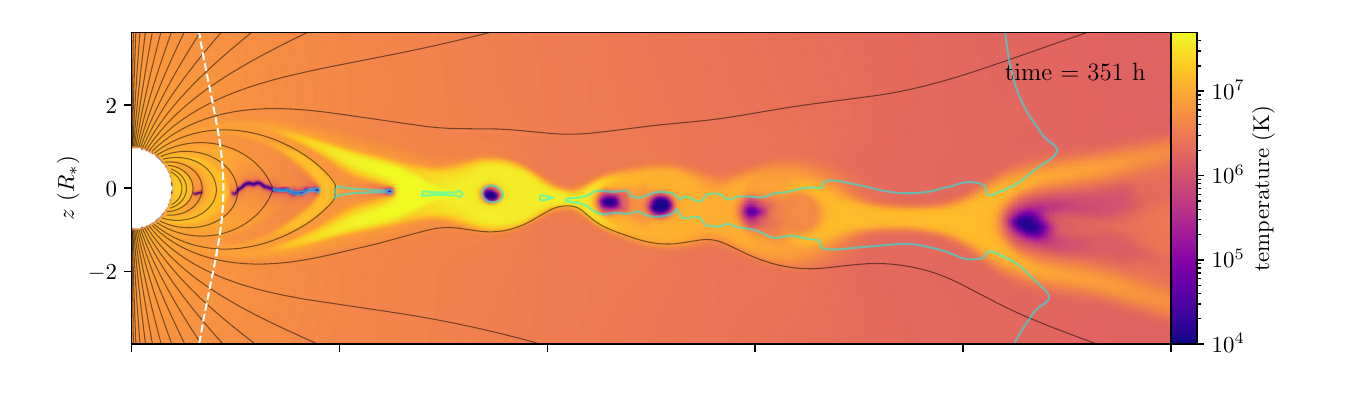}
    \includegraphics[width=\textwidth,trim={0.85cm, 0cm, 0.95cm, 0.3cm},clip]{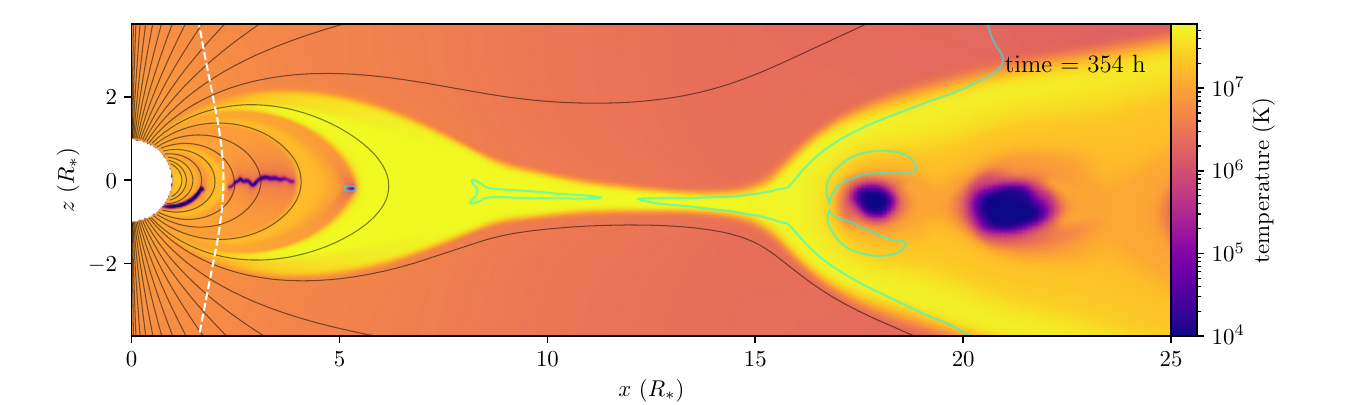}
	\caption[]{Time series of the 2D temperature profile for the centrifugal magnetosphere case. There are five distinct phases of the slingshot prominence process. From top to bottom: Evaporation raises the density of the mangetosphere. Second, the thermal instability leads to gas cooling to chromospheric temperatures within the closed magnetosphere, not just at $R_{\rm{K}}$. Third, cold gas below $R_{\rm{K}}$ drains to the surface and gas beyond $R_{\rm{K}}$ undergoes centrifugal breakout, elongating to a current sheet. Forth, the current sheet, unstable to the tearing instability, divides into a series of plasmoids. Finally, the plasmoids coalesce into larger structures as they pass beyond $R_{\rm{A}}$. The dotted white line indicates the Kepler co-rotation radius and the cyan solid line shows the Alfv\'{e}nic Mach surface. A movie version of this figure is available online.
 \label{fig:evolution_cnt}}
\end{figure*}

\begin{figure*}
	\centering
	\includegraphics[width=\textwidth,trim={0cm, 0cm, 0cm, 0cm},clip]{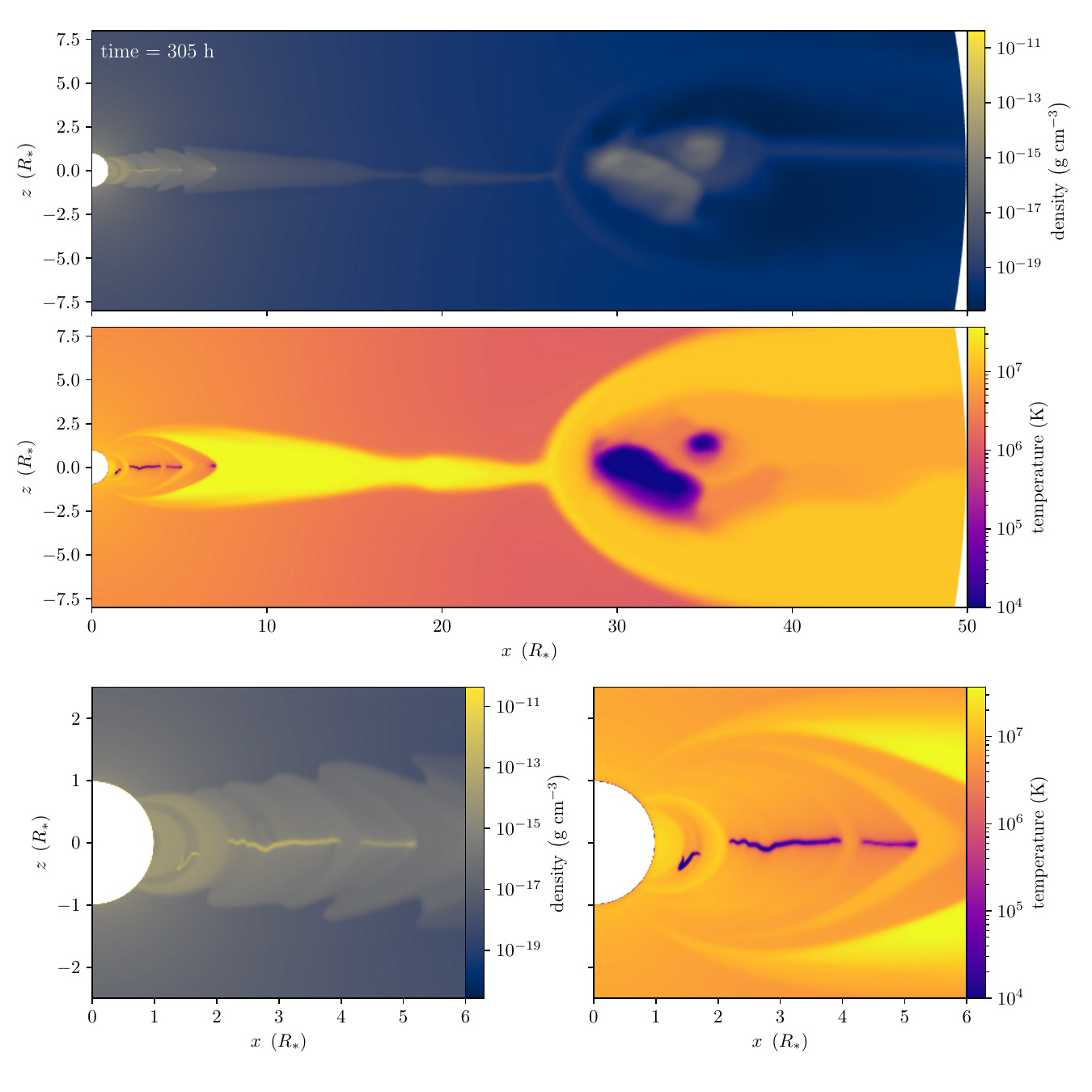}
	\caption[]{Example simulation snapshot showing density and temperature profiles at a time coinciding with the ejection of prominence material. The upper two panels show the large-scale overview with the ejected prominence at $\sim~30~R_{\ast}$. The bottom two panels show a zoomed-in portion of the inner magnetosphere. This region shows prominence material in three different stages: gas that is falling back to the stellar surface, that is suspended at co-rotation and that is undergoing centrifugal ejection. \label{fig:2d_profile}}
\end{figure*}

\begin{figure*}
	\centering
	\includegraphics[width=\textwidth,trim={0cm, 0cm, 0cm, 0cm},clip]{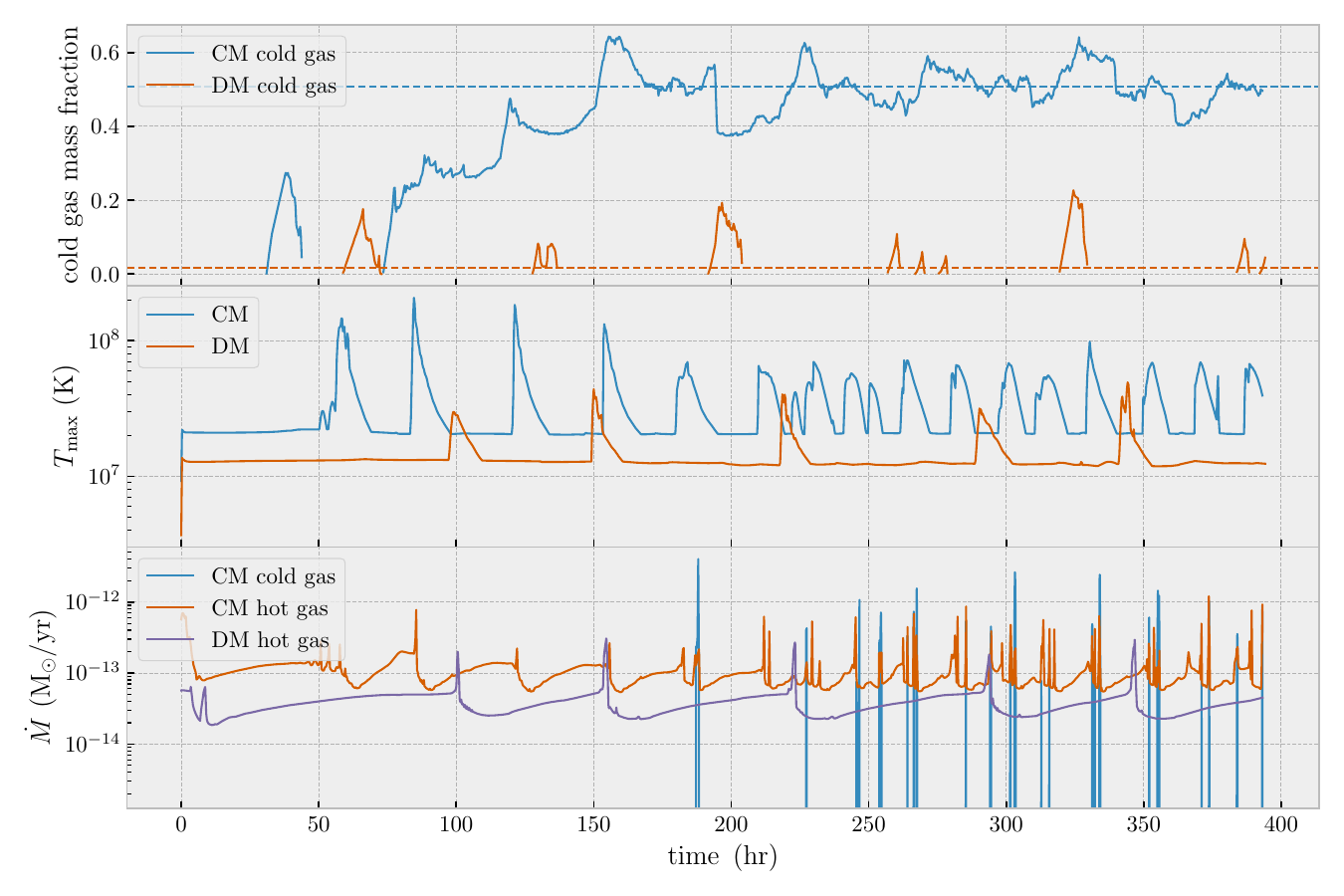}
	\caption[]{Top: time series of the cold gas mass fraction in the both simulation volumes. Here, as with the rest of the analysis, we define the boundary between hot and cold gas as 80000 K. Once a quasi-steady-state has been reached ($t~>~300$ hr), the cold gas component of the CM has reached a time average of 51\% (horizontal blue dashed line) and for the DM has reached 2\% (horizontal orange dashed line) of the total gas in the simulations (excluding the chromosphere). Middle: The maximum temperature in the two simulations. Both exhibit rapid temperature spikes followed by a decay back to coronal temperatures. The origin of these spikes is the thermal energy released through magnetic reconnection that accompanies the tearing of the current sheet illustrated in Fig. \ref{fig:evolution_cnt}. The difference in occurrence rates of these spikes is due to the rotational periods of the two simulated stars. Bottom: mass-loss rates for the two magnetosphere types and in the case of the centrifugal magnetosphere, the division between hot and cold gas. In steady state ($t~>~300$ hr), we find that mass-loss rate in cold gas is 21\% that of the total. We also find that the dynamical magnetosphere is losing mass an order of magnitude more slowly than the centrifugal case. \label{fig:time_series}}
\end{figure*}

\begin{figure}
	\centering
	\includegraphics[width=0.5\textwidth,trim={0cm, 0cm, 0cm, 0cm},clip]{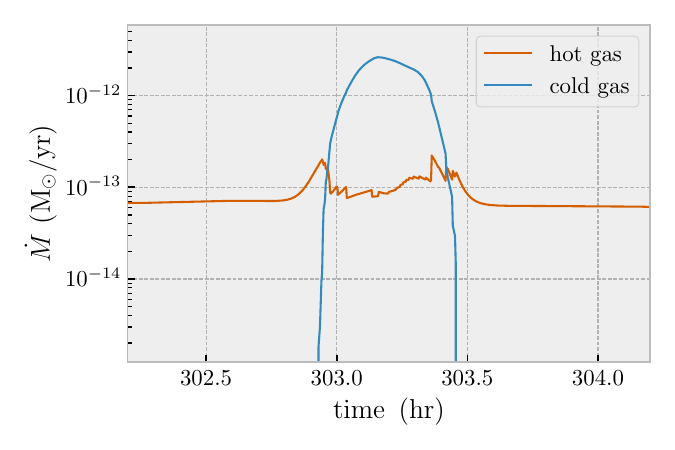}
	\caption[]{Mass-loss rate for a single slingshot event for both the background hot gas (orange line) and cold gas (blue line) from the CM. At its peak, the cold gas mass-loss is 35 times greater than the background hot gas mass-loss. \label{fig:mass-loss_single}}
\end{figure}

\begin{figure*}
	\centering
	\includegraphics[width=0.96\textwidth,trim={0cm, 0cm, 0cm, 0cm},clip]{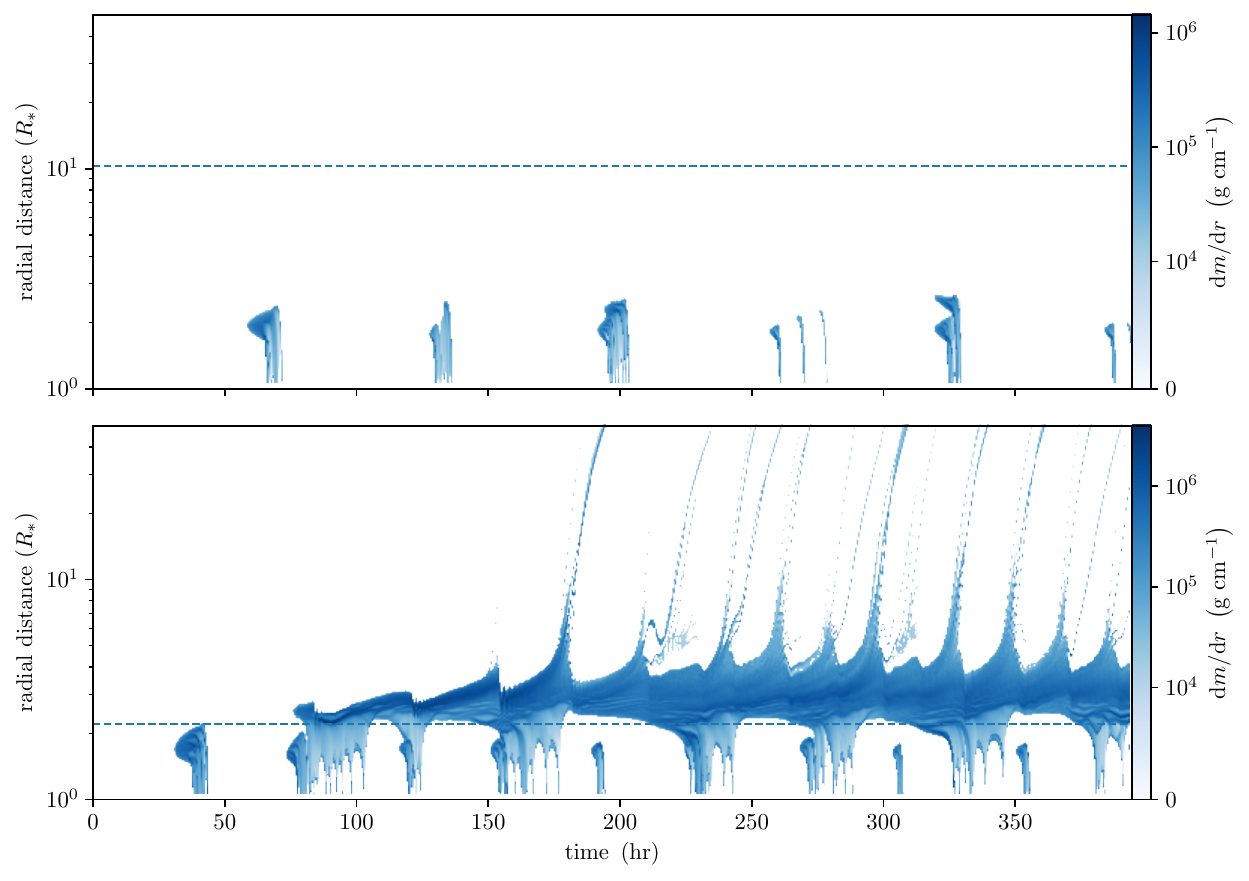}
	\caption[]{Time series of the radial mass distribution for only the cold gas component. Both plots show the equatorial co-rotational radius ($R_{\mathrm{K}}$) as the dashed line. Top: dynamical magnetosphere showing the formation of cold gas and its subsequent fall back to the stellar surface, coronal rain. Bottom: centrifugal magnetosphere showing an altogether more nuanced picture. The coronal rain seen in the dynamical magnetosphere is still present, but here it is only a subset of the overall behaviour. On top of this is a reservoir of cold gas which supplies gas for the periodic breakout of slingshot. These slingshots can be seen as upward moving tracks in the colour map above this reservoir. \label{fig:radial_mass}}
\end{figure*}

Here we will lay out the differences between the dynamical and centrifugal magnetospheres and demonstrate that it is the CM that generates slingshot prominences while the DM only generates coronal rain. As the results for the CM are more extensive, we will start with the DM results.

\subsubsection{Dynamical magnetosphere}
\label{sec:DM}

Fig. \ref{fig:evolution_dyn} illustrates three distinct stages of coronal rain formation. Initially, evaporation from the chromosphere increases the density in the closed magnetosphere. Once the density in the closed-field region is high enough, cooling by radiation leads to thermal runaway and gas cools to chromospheric temperatures ($10^{4}$ K). We see sympathetic cooling across field lines and cold gas extends from the apexes of the loops down to the chromosphere before finally the closed magnetosphere returns to only hot gas.
These stages are summarised as:
\begin{enumerate}
    \item The chromospheric evaporation stage.
    \item The cooling stage.
    \item The condensation and evacuation stage.
\end{enumerate}
Once established, these stages repeat in a cyclic manner for the rest of the simulation. This process is identical to solar coronal rain \citep{Antolin2020}. While the formation of stellar coronal rain is observable and an indicator of magnetic activity, it does not impact the properties of the extended wind, for example mass-loss rate, and therefore plays no role in the long term evolution of the star.

\subsubsection{Centrifugal magnetosphere}

The formation phases of the slingshot prominence process are shown in Fig. \ref{fig:evolution_cnt}. First, as in the DM case, evaporation from the chromosphere increases the density in the closed magnetosphere. Second, the thermal instability leads to gas cooling to chromospheric temperatures within the closed magnetosphere, both above and below $R_{\rm{K}}$. 

Third, cold gas below $R_{\rm{K}}$ drains to the surface forming coronal rain, while cold gas beyond $R_{\rm{K}}$ undergoes centrifugal breakout, elongating to a thin sheet as it leaves the closed magnetosphere. This thin sheet remains intact as it passes through the magnetic \textit{Y-null} point at the apex of the helmet streamer. This is important as it demonstrates that the timescale for breakout of a slingshot (and therefore the time between subsequent breakouts) is not simply the timescale of the tearing instability but rather the time taken for the cold gas pressure to exceed the magnetic tension. This is a function of the rate of evaporation of gas form the chromosphere and therefore the feeding rate of the stable region above $R_{\rm{K}}$. 

Fourth, the current sheet, unstable to the tearing instability, divides into a series of plasmoids. Finally, the plasmoids coalesce into larger structures as they pass beyond $R_{\rm{A}}$. 

In Fig. \ref{fig:evolution_cnt}, the dotted white line indicates the Kepler co-rotation radius and the solid cyan line shows the Alfv\'{e}nic Mach surface. Crucially, $R_{\rm{K}}$ is inside $R_{\rm{A}}$, signifying a CM, in contrast to Fig. \ref{fig:evolution_dyn} where $R_{\rm{K}}$ is either at or outside $R_{\rm{A}}$. The five stages of the CM are therefore:

\begin{enumerate}
    \item The chromospheric evaporation stage.
    \item The cooling stage.
    \item Condensation and draining {\it below}, and centrifugal breakout {\it above} $R_{\rm{K}}$.
    \item The tearing instability produces plasmoids.
    \item The plasmoids coalescence beyond $R_{\rm{A}}$.
\end{enumerate}

A major departure in the CM case from the DM case is that there is no stage where the magnetosphere is wholly evacuated of cold gas. Low lying loops, well below $R_{\mathrm{K}}$, do exhibit this, but between $R_{\mathrm{k}}$ and $R_{\mathrm{A}}$ cold gas is constantly forming and draining. {See Fig. \ref{fig:2d_profile} for not only the temperature but also the density profiles including a close up of both. We have concentrated on the temperature and density as these quantities completely define the presence of condensations. Please see Appendix \ref{sec:app_vel} for plots of the velocity components}. We will see in the following section that this cold gas component forms a significant fraction of the total magnetosphere mass.

\subsection{Time series}

Below we analyse the time dependent quantities that show how the multi-phase gas evolves over the simulation (which last for a total of 393 hr). This maximum time was chosen to strike a balance between simulating a sufficient number of slingshots and the computational resources used. We concluded that 393 hr allowed us to draw firm conclusions and to perform a limited statistical comparison to observations. Fig. \ref{fig:stats} shows our time series results.

\subsubsection{Mass balance}

The top plot of Fig. \ref{fig:time_series} shows a series of the cold gas mass fraction in both simulation volumes. Here, as with the rest of the analysis, we define the boundary between hot and cold gas as 80000 K. For the CM, once a quasi-steady-state has been reached, the cold gas component is 51\% of the total gas in the simulation (excluding the chromosphere) and remains at approximately this value for the rest of the simulation. For the DM, the cold gas component is 2\%, a much smaller fraction of the magnetosphere. It is also intermittent, going to zero during the DM stage (i) as described in Section \ref{sec:DM}.

\subsubsection{Maximum temperature as an observational proxy}

In the middle plot of Fig. \ref{fig:time_series} we show the time series of the maximum temperature ($T_{\mathrm{max}}$) in the simulation. This is determined by finding $T_{\mathrm{max}}$ on the entire numerical grid at each time step. The location of $T_{\mathrm{max}}$ is always in the region of the reconnected field, upstream of the recently ejected prominence. Each successive breakout is accompanied by a spike ($T_{\mathrm{max}}\lesssim~10^{8}$ K) which is up to an order of magnitude above the background coronal temperature ($\sim~2\times10^{7}$ K). This behaviour is also seen in the DM simulation (though with a smaller $T_{\mathrm{max}}$), which has no centrifugal breakout, so this cannot be attributed to the slingshots. {In both the CM and DM case, between reconnection events, the magnetosphere has to relax back to equilibrium. This is evidenced in Fig. \ref{fig:evolution_dyn} where the field lines are stretched out over the frames shown.}

At $\sim~30$ hr after the condensations have formed (either into coronal rain in the DM case or slingshots in the CM case), there is a pinching of the helmet streamer and reconnection of the magnetic field. This accompanies the spike in $T_{\mathrm{max}}$, which may have a signature in observations. Stellar flares are observed in UV, x-ray and radio wavelengths \citep{Benz2010ARAA} and are characterised by a sudden rise in emission and a decay to background levels. 

If we contrast the DM to the CM, we see a lower background temperature, less frequent spikes and lower $T_{\mathrm{max}}$ spikes. 

{The amplitude of our heating model in our simulations is proportional to the magnetic field strength. The CM has $B_{0}\sim~800$ G and the DM $B_{0}\sim~200$ G dipole moments. Therefore the CM is necessarily hotter as there is more energy being deposited into the simulation than in the DM case. This explains why the baseline $T_{\mathrm{max}}$ is lager for the CM. The spikes are due to reconnection, with a larger field strength in the CM case, there is more magnetic energy released by reconnection than the DM case, leading to a lager $T_{\mathrm{max}}$ spike for the CM. The entire helmet streamer region is subsonic, therefore shock heating is not responsible for the high temperature, contrary to the massive star case where the entire wind is supersonic almost from the surface of the star.} This is consistent with the observational result that flare activity is correlated with rotation rate for young stellar objects. The faster a star spins, the more it flares \citep{Gunther2020A, Vida2024arXiv}. 

Calculating synthetic observations from our simulation result would give us a direct comparison to observations, this will be the subject of a dedicated future study. 

\subsubsection{Mass-loss rates}

The mass-loss rates are measured by integrating the radial component of the momentum over a surface at $r~=~25R_{\ast}$. The results are shown in the bottom plot of Fig. \ref{fig:time_series} for both the CM and DM. We show both the cold ($\dot{M}_{\mathrm{c}}$) and hot ($\dot{M}_{\mathrm{h}}$) components in the CM, but only the hot component for the DM case, as no cold gas escapes the the star. Another difference between the two magnetosphere types is the magnitude of $\dot{M}_{\mathrm{h}}$. Our DM has mass-loss rates that are always lower by approximately an order of magnitude. This is consistent with the difference in heating rates used in the two simulations (see Table \ref{tab:parameters}). The faster rotator has more heating, giving greater chromospheric evaporation, leading to a denser magnetosphere and wind, resulting in the higher mass-loss rate.

Turning to $\dot{M}_{\mathrm{c}}$, for times $t~>~200$ hr, the simulation starts to generate slingshots. The mass flux associated with each slingshot is plotted as the blue spikes overlaying the background $\dot{M}_{\mathrm{h}}$. By integrating the area under these curves we find that the accumulated mass-loss in cold gas is 21\% of the total mass-loss, a significant fraction.

To better understand the mass-loss associated with an individual slingshot, we plot a single breakout event in Fig. \ref{fig:mass-loss_single}. During the event, $\dot{M}_{\mathrm{c}}$ is 35 times larger than $\dot{M}_{\mathrm{h}}$. If we recall that the mass-loss calculation is carried out over a surface containing the star at $r~=~25R_{\ast}$ and for the hot gas includes the fast and slow wind at all latitudes, this means that for the period of a slingshot, the star's mass-loss is dominated by cold gas.

\subsubsection{Low- and high-lying loop solutions}

Cold gas in the magnetosphere from either the DM or CM is highly dynamic. Simple formation and draining occurs cyclically for the DM but in the CM case, cold gas drains, escapes outwards or is suspended all at the same time. To better understand the nature, intermittent or otherwise, of each of these states we plot the radial mass distribution of both simulations in Fig. \ref{fig:radial_mass} in the manner of \cite{ud-Doula2008} and \cite{Daley-Yates2019}.

The DM shows condensation and draining in the form of coronal rain. The CM shows strikingly different behaviour. After an initial coronal rain event, a reservoir forms above $R_{\mathrm{K}}$ and remains for the length of the simulation. Coronal rain still occurs in a similar manner to the DM. This can be seen below the reservoir of stable material. There is also draining from the bottom of the reservoir to the stellar surface, similar to coronal rain. Above the reservoir cold gas breaks out of the closed magnetosphere, indicating slingshot events. These form upward trending tracks from the reservoir to the edge of the simulation domain\footnote{The apparent dotted nature of the tracks in Fig. \ref{fig:radial_mass} is due to the cadence of the simulation outputs which is every 1000 s (with inflate cadence the tracks would be continuous)}. The gradient of these tracks indicate the velocity of the slingshots as they leave the magnetosphere. We do not attempt to quantify the velocities from Fig. \ref{fig:radial_mass} as we measure this directly from simulation outputs and present it in the next section. We note however that the gradient of the tracks decreases beyond $>30 \ R_{\ast}$, indicating that the slingshots begin to slow down as they leave the magnetosphere. 

Fig. \ref{fig:radial_mass} illustrates the two types of loops that form in the magnetosphere of a young, fast rotating star. Low-lying loops support the formation of coronal rain. We have seen and reported the low-lying loop solutions in our previous paper \cite{Daley-Yates2023}. Indeed low-lying loop solutions are seen on the Sun as Solar coronal loops that exhibit Solar coronal rain. The second, high-lying loops, support the formation of slingshots and are not seen on the Sun. The division between these two solutions depends on the position of the co-rotation radius. $R_{\mathrm{K}}$ must coincide with closed magnetic loops capable of forming condensations. Fig. \ref{fig:radial_mass} shows the position of $R_{\mathrm{K}}$ for both magnetospheres. The presence of high-lying loop solutions does not preclude the existence of low-lying loops. In the CM we see both families of solutions. These two loop solutions were predicted by the analytic work of \cite{jardine91,waugh2019, waugh2021, waugh2022}. We report here numerical conformation of this idea.

\subsection{Slingshot statistics}
\label{sec:results_stats}

\begin{figure}
	\centering
	\includegraphics[width=0.5\textwidth,trim={0cm, 0cm, 0cm, 0cm},clip]{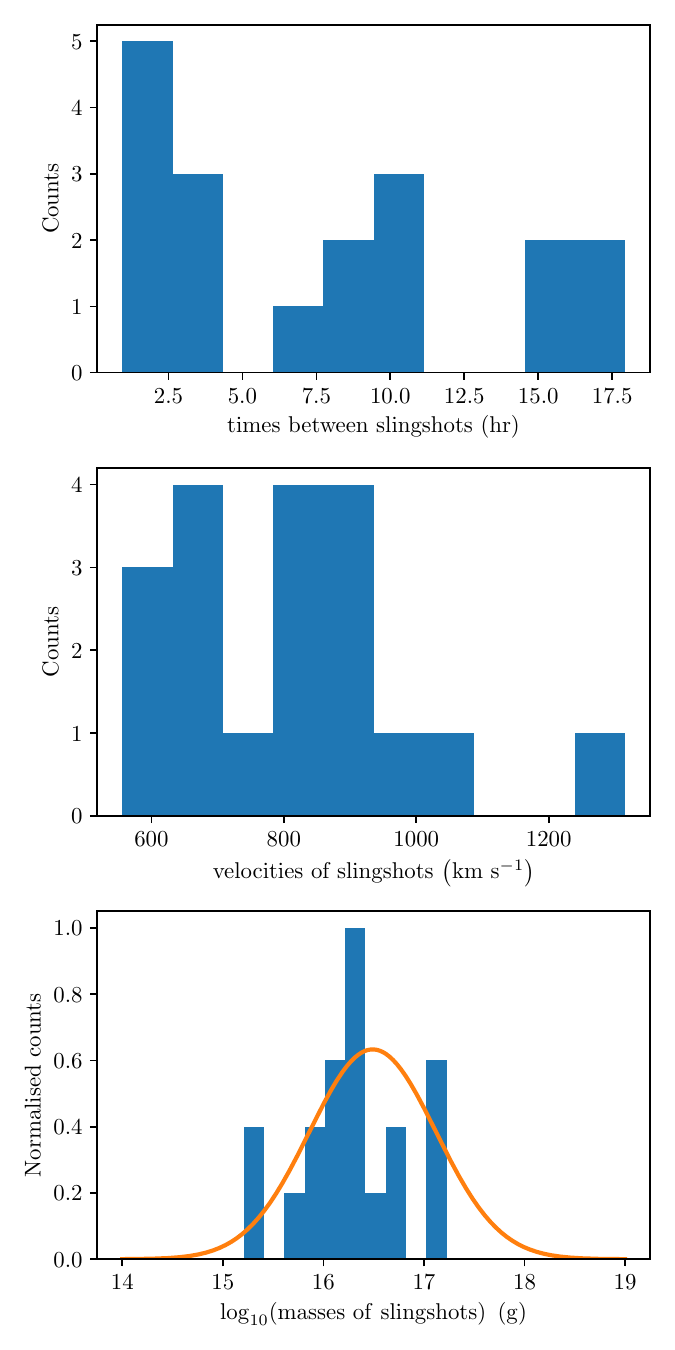}
	\caption[]{Statistical distribution of the 18 separate slingshot prominences seen in our CM simulation. From top to bottom: time between breakouts, line of sight velocities of each breakout and finally the masses of the breakouts. This bottom plot also a fit to the observational results reported by \cite{Vida2019}. In order to make the mass results comparable to the observational statistics, we have normalised the counts. \label{fig:stats}}
\end{figure}

Here we investigate the statistical properties of the CM simulation. In total there are 18 slingshots recorded. The time between slingshots varies from 0.9 - 17.9 hrs (approximately $0.075 - 1.5$ stellar rotations) with an average of 7.7 hrs. Breakout times around 2.5 hr are due to multiple blobs of gas being ejected as part of the same breakout event, and are in fact one slingshot. We do not count these as separate events since the blobs can and do coalesce at larger radii. For the 7.5 - 17.5 hrs range it is perhaps useful to compare to the observations of the K3 dwarf star Speedy Mic (BO Mic, HD 197890) by \cite{Dunstone2006} who measured stable prominences lasting for more than 13 stellar rotations. They found prominence structures at an average height of $2.85\pm0.54 R_{\ast}$, a very similar height to our stable reservoir situated at $>2.21 R_{\ast}$ in our CM simulation (see Fig. \ref{fig:radial_mass}). We argue that what \cite{Dunstone2006} observed was not cold gas making up slingshot prominence, but the stable reservoir which supplies the cold gas to the slingshots. The times between our slingshots is also much shorter than the 13 rotations that they report. This is consistent with the idea that this is in fact the timescale for the reservoir to evolve and not the time taken for individual prominences to breakout as slingshots.

Our slingshots have velocities between 556 - 1315 km/s with an average of 813 km/s. The velocities do not agree with \cite{Vida2019} whose values for the blue shifted velocities are between 0 - 800 km/s. In our case the slingshots are traveling along the line of sight, so all samples are the fastest possible. For the observations, the lines of sight velocities are components and not the total velocities, making our velocity distribution an upper limit on the observations. This is what we see when we compare them to Fig. 7 of \cite{Vida2019}. We have one outlier in the velocity beyond the results of \cite{Vida2019} with a value of 1315 km/s.

\cite{Inoue2023} reported observing the very active RS CVn-type binary V1355 Orionis releasing a superflare of $7 \times 10^{35}$ ergs along with a feature showing an H$_{\mathrm{\alpha}}$ excess travelling between 760 - 1690 km/s with a mass between $9.5 \times 10^{18}$ g - $1.4 \times 10^{21}$ g. While this velocity range includes our outlier, the mass range far exceeds any of the slingshots we measure in our simulation which are between $1.61\times10^{15}$ - $1.68\times10^{17}$ g with an average of $3.95\times10^{16}$ g. The K-type subgiant component of V1355 Orionis is a larger star than those we have simulated and therefore has different values for $R_{\mathrm{K}}$ and $R_{\mathrm{A}}$ and may support larger prominences. It has a co-rotation radius of $1.8 R_{\ast}$ above its surface, making it comparable in this respect to our simulated CM star. However for the slingshot mechanism to be responsible for the superflare on V1355 Orionis, its magnetic field would have to be sufficient to give it a centrifugal magnetosphere. Unfortunately we do not have any information on the magnetic field strength or structure of V1355 Orionis, so our speculation ends here and can only state that the velocity range of our simulated slingshots does coincide with the velocity associated with this superflare.

Our mass range of $1.61\times10^{15}$ - $1.68\times10^{17}$ g for our slingshots agrees with the statistics of \cite{Vida2019} remarkably well. To underline this, we plot the normal fit to their data that they report over our statistical sample. Both quantities are normalised for ease of comparison. There is a tight agreement between observation and our results. \cite{Dunstone2006b} reported masses for the largest prominences of Speedy Mic to be $0.5\times10^{17}$ - $2.3\times10^{17}$ g, which agrees well with our results. Both the statistical sample of \cite{Vida2019} and this single stellar observation agree much better with our simulated mass range of $1.61\times10^{15}$ - $1.68\times10^{17}$ g than the superflare observation of V1355 Orionis, despite the velocities being so different. We stress our results are not directly comparable to that of \cite{Vida2019}, as each data point in their observations is a snapshot of the whole magnetosphere of an individual star. Our statistics are made up of only the slingshots as measured when they pass through $r~=~25R_{\ast}$, and we ignore any other gas in the simulation, cold or otherwise. As such we regard our results as a possible subset, embedded in the complete results presented by \cite{Vida2019}. We also only look at their blue-shifted results as this is relevant to our slingshots.

\section{Conclusions}

We have performed numerical simulations of stellar coronae that demonstrate the formation of massive clumps of cool gas condensing out of the hot corona. The {\it onset} of condensation depends on the heating process. An increase in coronal heating evaporates chromospheric material into the corona, raising the density sufficiently high for the onset of thermal collapse. The {\it subsequent dynamics} of these clumps, however, depend on the star's rotation rate. 

We find two regimes that can be classified by the ratio of the co-rotation radius $R_{\mathrm{K}}$ to the Alfv\'en radius $ R_{\mathrm{A}}$. This centrifugal-dynamical magnetosphere framework was first developed in the massive stars community by \cite{ud-Doula2008} and \cite{Petit2013} and expanded to cool stars by \cite{Villarreal2017}. The more slowly-rotating of our two stars is in the {\it dynamical regime}  (DR: $R_{\mathrm{K}} < R_{\mathrm{A}}$). In this case, condensations form below the co-rotation radius and fall back towards the stellar surface. We can identify this pattern of downflows as the stellar equivalent of solar coronal rain. This behaviour is also seen in the more rapidly-rotating of our two stars, which is in the {\it centrifugal regime} (CR: $R_{\mathrm{K}} > R_{\mathrm{A}}$). In this case, condensations not only form below the co-rotation radius but also at and above it.  At the co-rotation radius they grow to form a quasi-stable mass reservoir, from which there are regular centrifugal breakout events as the mass grows beyond the point of magnetic confinement and some of it is ejected from the star. We identify the large stable reservoir with stellar ``slingshot prominences'' and the ejecta with the fast-moving absorption features with which they are associated.

In the slowly-rotating dynamical magnetosphere case, the hot wind carries away mass at a rate $3.6\times10^{-14} \ \mathrm{M}_{\odot}/\mathrm{yr}$ which is a little greater than the present-day Sun. There is no mass-loss from the condensations. In the faster-rotating centrifugal magnetosphere case, the hot gas removes mass more rapidly, at a rate of $9.4\times10^{-14} \ \mathrm{M}_{\odot}/\mathrm{yr}$. In addition, however, the cold gas ejected from the large reservoir at co-rotation radius removes mass at a rate of $2.7\times10^{-14} \ \mathrm{M}_{\odot}/\mathrm{yr}$. This comprises 21\% of the total mass lost. Indeed, some 51\% of the coronal mass is in cold gas. This is potentially a significant contribution that is unaccounted for in models of the hot wind.

The distribution of clump masses clusters around $10^{16}$ g and the line of sight velocities range between $600 - 1000$ km/s, with a single outlier $>1200$ km/s. These results agree well with the observational statistics of \cite{Vida2019} for clump masses, but as discussed in Section \ref{sec:results_stats}, our simulated velocities agree with the upper limit of the observations.

Both regimes display well-defined periodicities. In the slowly-rotating dynamical magnetosphere case, there are periodic down-flows where the cold gas drains completely from the corona. The interval between these events is $\sim~75$ hr. In the faster-rotating centrifugal case, the region below the co-rotation radius also shows regular draining events, both from clumps that form below the co-rotation radius, and also from the large mass reservoir at the co-rotation radius itself. The part of the mass reservoir that lies about the co-rotation radius also loses mass periodically. These centrifugal breakout events have periods from 7.5 - 17.5 hr (there is a shorter period of 2.5 hr due to fragmentation into sub-clumps). 

{Both our simulation have been limited to 2D, in future studies we intend to move to 3D. This increase in dimensionality will allow us to quantify such things as: the number of prominence structures a star can support, weather they form a disk or discrete structures and allow us investigate observational signatures with a strong geometric component, such as H$_{\alpha}$ tracks.}

Contrasting the two magnetosphere types that we have simulated shows that there are two distinct types of solutions, high lying and low lying loops. Both are seen in the centrifugal magnetosphere but only the low lying loops are seen in the dynamic magnetosphere. Low lying loops only produce the stellar equivalent of solar coronal rain whereas high lying loops produce not only rain but also what has been dubbed ``slingshot prominences''. The modern day Sun only exhibits the low lying loops, but in the past when it was rotating faster it would have had both low and high loop types. This means that the early solar system planets would have evolved under the influence of slingshots. As a result we propose the slingshot mechanism as a new type of space weather for young stellar systems.

\section*{Data availability}
A CSV file of the data presented in Fig. \ref{fig:stats} is provided online.

\section*{Acknowledgements}

The authors thank the reviewer for their helpful comments and suggestions; which improved the quality and content of the publication. SD-Y and MJ acknowledge support from STFC consolidated grant number ST/R000824/1. This work was performed using the DiRAC Data Intensive service at Leicester, operated by the University of Leicester IT Services, which forms part of the STFC DiRAC HPC Facility (www.dirac.ac.uk). The equipment was funded by BEIS capital funding via STFC capital grants ST/K000373/1 and ST/R002363/1 and STFC DiRAC Operations grant ST/R001014/1. DiRAC is part of the National e-Infrastructure. This research was supported by the International Space Science Institute (ISSI) in Bern, through ISSI International Team project 545 (``Observe Local Think Global: What Solar Observations can Teach us about Multiphase Plasmas across Physical Scales''). For the purpose of open access, the authors have applied a Creative Commons Attribution (CC BY) licence to any Author Accepted Manuscript version arising.

\section*{ORICD iDs}

Simon Daley-Yates \orcidlink{0000-0002-0461-3029} \url{https://orcid.org/0000-0002-0461-3029} \\
Moira M. Jardine \orcidlink{0000-0002-1466-5236} \url{https://orcid.org/0000-0002-1466-5236} \\

\bibliographystyle{mnras}
\bibliography{references}

\appendix

\section{Angular components}
\label{sec:app_vel}
\begin{figure}
	\centering
	\includegraphics[width=0.49\textwidth,trim={1cm, 0cm, 0cm, 1cm},clip]{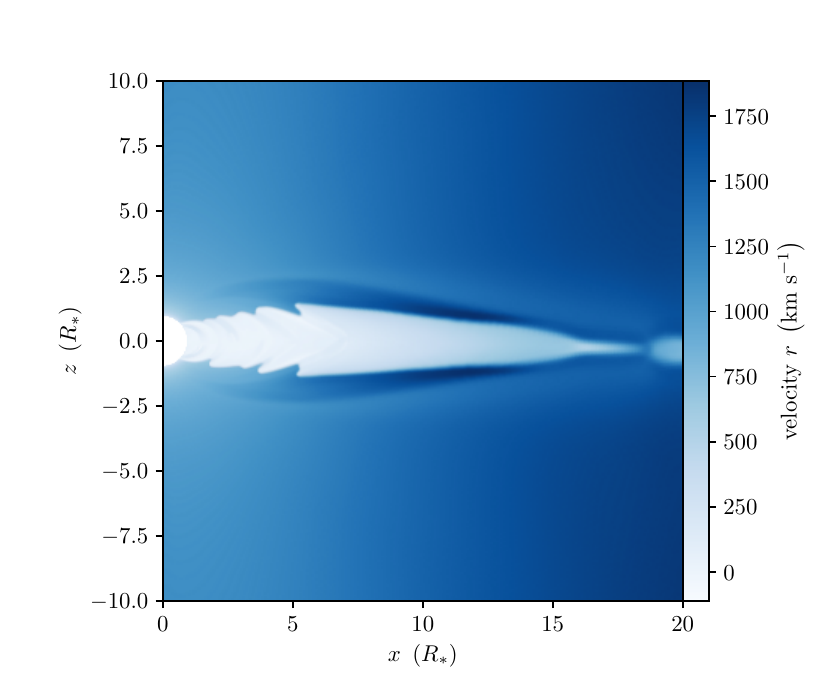}
    \includegraphics[width=0.49\textwidth,trim={1cm, 0cm, 0cm, 1cm},clip]{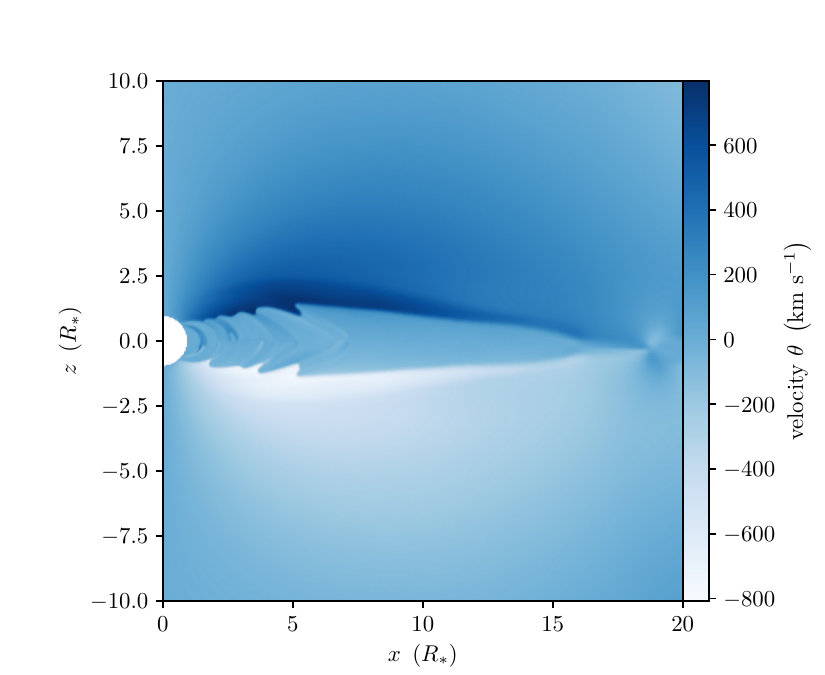}
 	\includegraphics[width=0.49\textwidth,trim={1cm, 0cm, 0cm, 1cm},clip]{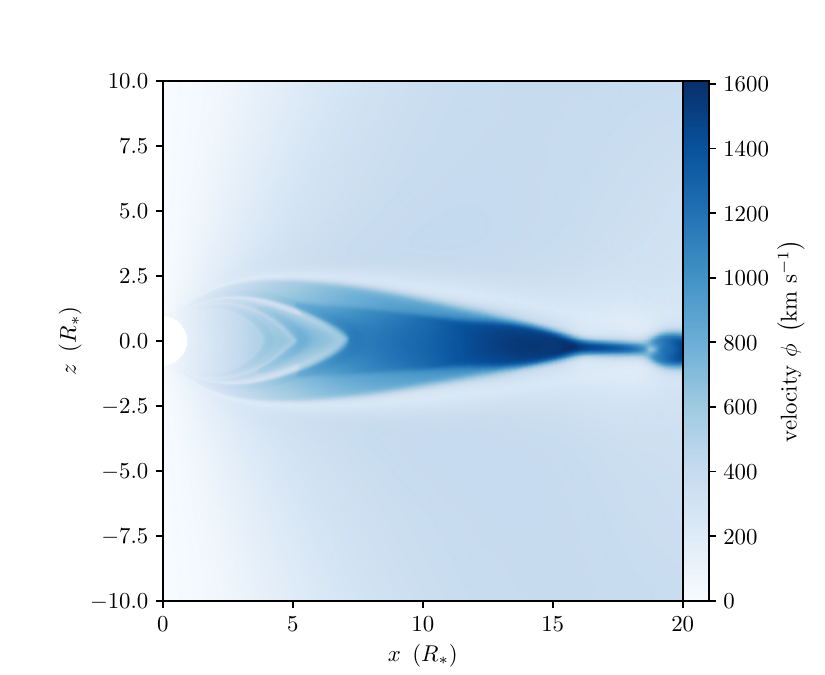}
	\caption[]{Radial (top), poloidal (middle) and azimuthal (bottom) velocity in the observers reference frame, at 305 hr from the start of the simulation.
    \label{fig:vt_vp}}
\end{figure}

Figure \ref{fig:vt_vp} shows the velocity components.


\label{lastpage}

\end{document}